\pdfoutput=1
\documentclass[aps,prb,onecolumn,groupedaddress]{revtex4}

\usepackage{float}
\usepackage{amsmath}
\usepackage{amssymb}
\usepackage{siunitx}
\usepackage{chemformula}
\usepackage{graphicx}
\usepackage{dcolumn}
\usepackage{bm}
\usepackage{siunitx}
\usepackage{lmodern}

\usepackage[T1]{fontenc}
\usepackage[utf8]{inputenc}
\usepackage{textcomp}
\usepackage{epstopdf}
\usepackage{natbib} 
\usepackage{hyperref}
\usepackage{ulem}
\usepackage{soul,color}
\makeatletter
\begin{document}

\title{Radiation damage effects on Helium diffusion in zircon}
\author{A. Diver$^1$, O. Dicks$^1$, A. M. Elena$^2$, I. T. Todorov$^2$, T. Geisler$^3$, K. Trachenko$^1$}
\address{$^1$ School of Physics and Astronomy, Queen Mary University of London, Mile End Road, London, E1 4NS, UK}
\address{$^2$Daresbury Laboratory STFC UKRI, Scientific Computing Department, Keckwick Lane, Daresbury WA4 4AD, Cheshire, UK}
\address{$^3$ Institut f\"{u}r Geowissenschaften, Universit\"{a}t Bonn, 53115 Bonn, Germany}

\begin{abstract}

We report the effects of radiation damage on helium diffusion in zircon using data from molecular dynamics simulations. We observe an increase in activation energy for helium diffusion as a result of radiation damage and increasing structural disorder. The activation energy in a heavily damaged region is smaller than in a completely amorphous system which is  correlated with remaining order in the cation sublattices of the damaged structure not present in the fully amorphized system. The reduction of activation energy is related to the disappearance of fast diffusion pathways that present in the crystal. Consistent with the change in activation energy, we observe the accumulation of helium atoms in the damaged structure and discuss the implications of this effect for the formation of helium bubbles and zircon's performance as an encapsulation material for nuclear waste.
\end{abstract}

\maketitle

\section{Introduction}

The effects of radiation damage on zircon are one of the most well studied and characterized of any material. For many years it has been considered a prime candidate for the use as an encapsulation matrix for radioactive waste\cite{ewing_lutze_weber_1995}. The initial interest was excited when natural zircon samples were found to contain low concentrations of radioactive actinides and as such was considered a 'natural encapsulation matrix'. This has resulted in many years of experimental studies on natural zircon samples examining the direct impact of radiation damage via ion beam irradiation \cite{ewing1994radiation,murakami1991alpha,weber1990radiation,weber1991self}. These experiments observed that irradiation results in the amorphization of zircon, which is accompanied by a large volume expansion of approximately 18-20 \% \cite{weber1994radiation}. In more recent years the use of Molecular Dynamics (MD) simulations have been carried out to aid the understanding of radiation damage in zircon, predicting the formation of Si-O bridges, the percolation of damage, and density inhomogenities formed as a  direct result of simulated collision cascades \cite{Trachenko_2004per,PhysRevB.65.180102,diver2020evolution}. 

Crucial for the development of future safe wasteforms is determining the effect of helium buildup produced as a result of the alpha decays of the radioactive actinides present. Helium buildup in wasteforms has been linked to inducing embrittlement of the wasteform, swelling from helium bubble formation and potential pressurization of metal canisters enclosing the waste form \cite{weber1983review,weber1998radiation}. Decreased brittleness and microcracking as a result of swelling can lead to increased dissolution rates and influence the movement of actinides \cite{weber1998radiation}. These concerns are important due to helium bubbles being identified in different highly damaged natural minerals \cite{headley1981amorphous,lumpkin1986alpha,seydoux2016trapping}. 

Another result of the radioactive elements present in natural zircon is that it is often used as a thermochronometer, due to zircon (U–Th)/He (ZHe) decay \cite{reiners2005zircon,reiners2004zircon}. Thermochronometery gives us information on the history of the sample, its age and most importantly the rate at which the specific sample has been cooled. However for accurate ages to be calculated, the mechanisms of helium loss need to be well understood, as the age is calculated by the ratio of uranium and thorium to helium and lead. 
The alpha-decay of actinides in zircon introduces the helium (at high energies) whilst simultaneously changing the local structural environment effecting the rate of helium diffusion as a result of radiation damage. The destruction of crystallographic directions due to even low levels of radiation damage can decrease diffusivity by destroying fast diffusion pathways \cite{bengtson2012he,farley2007he}. This is due to the highly anisotropic nature of He diffusion in zircon. The most favourable pathway for He diffusion is along the [001] direction parallel to the \textit{\textbf{c}} axis \cite{guenthner2013helium,cherniak2009diffusion,reiners2004zircon,reiners2005zircon}. Even small changes in structure as the substitution of rare earth elements instead of Zr and Si produce differences in diffusion rates between natural samples and synthetic analogues \cite{farley2007he}. The production of voids induced by radiation damage \cite{danivsik2017seeing} can result in regions of up to 3 times greater concentration of helium than the surrounding undamaged crystal. This suggests that the rate of diffusion in a damaged, semi-amorphised region of zircon will be lower than that in a crystalline region.

Similar results are found when comparing the effect of radiation dose on diffusion in different samples \cite{guenthner2013helium}. Dose is often used to quantify the amount of 'damage' in a sample and is measured as the number of alpha-decays/g. As dose increases up to \num{1e16} $\alpha$/g the activation energy needed for diffusion also increases. However, further increasing the dose to \num{1e19} $\alpha$/g results in a decrease in the activation energy \cite{guenthner2013helium}. This indicates that the initial damage to the sample decreases diffusivity until a critical damage point is reached, after which further damage only increases diffusivity. This has been linked to the effect of percolation of damage creating larger regions of overlapped \cite{gautheron2020multi} damage, allowing easier He diffusion to take place along these pathways. In minimally damaged natural zircon samples, with doses of \num{1e15} $\alpha$/g, the mean distance between damaged regions can be as little as 5 \textmu m. This means all fast diffusion pathways are interrupted, or completely damaged, thus resulting in a regime where He atoms diffuse alternately between isolated damaged and crystalline regions\cite{ketcham2013geometric}. Taking this into account it can be difficult to compare diffusion data between different samples, as they all have different ages, doses and thermal histories.

Various computational modelling techniques have been used to study helium diffusion in zircon. Common practise is to use density functional theory (DFT) and in particular nudged elastic band to calculate the activation energy for helium diffusion from one minima to another.\cite{bengtson2012he,reich2007low,saadoune2009,gautheron2020multi}. The values for the activation energy vary depending on the parameters used, but in general  they agree with each other and experimental work that diffusion parallel to \textit{\textbf{c}} is more favourable than any other direction. It is also worth noting that calculations of activation energies for helium diffusion are much smaller than experimental studies, due to simulations being conducted on perfect crystals with no elemental impurities, defects or radiation damage which can cut off the c-channel along which He prefers to diffuse. The only computational work we know taking into account the effect of radiation damage on helium diffusion calculates diffusion based on the hopping of helium atoms between interstitial sites \cite{gautheron2020multi}, and models the damage by blocking some percentage of the sites depending on the degree of damage.

The aim of this work is to use molecular dynamics (MD) to calculate activation energies for helium diffusion in three different zircon structures; a perfect crystal, a homogeneous amorphous structure and a damaged crystalline structure produced as a result of 6 overlapping collision cascades of a 70 keV U ion. This is the first time such calculations have be performed using large scale molecular dynamics. We will present Arennhius plots of the different structures at a range of temperatures and show that radiation damage results in an increase in activation energy for helium diffusion. We will also show that even a small amount of retained order can lead to a much faster rate of diffusion than a completely amorphous structure. We also show that there is a greater build up of helium in damaged regions over time, and discuss the implications of this effect for helium bubbles and waste form performance.

\section{Methods}

The molecular dynamics (MD) package \texttt{DL\char`_POLY\char`_4} (version 4.10 with changes released in version 5.0.0)
  \cite{todorov_smith_trachenko_dove_2006} has been used to model He diffusion in zircon systems, due to its ability to simulate large systems with a large number of atoms, in parallel, by using domain decomposition.

The interatomic potentials used in this work to model zircon have been described in previous papers \cite{Trachenko_2004per,diver2020evolution}. The Zr-O and O-O pairwise potentials are in the form of a Buckingham potential, with the Si-O pairwise potential taking the form of a Morse potential. The interaction between He and the constituent atoms in zircon must also be included, and are taken from previous work \cite{reich2007low}. We also include a term to model the He-He interaction which has been well used and tested in MD simulations of uranium dioxide \cite{grimes1990behaviour,govers2009molecular}.

\begin{figure}
\centering
\includegraphics[width = 8cm]{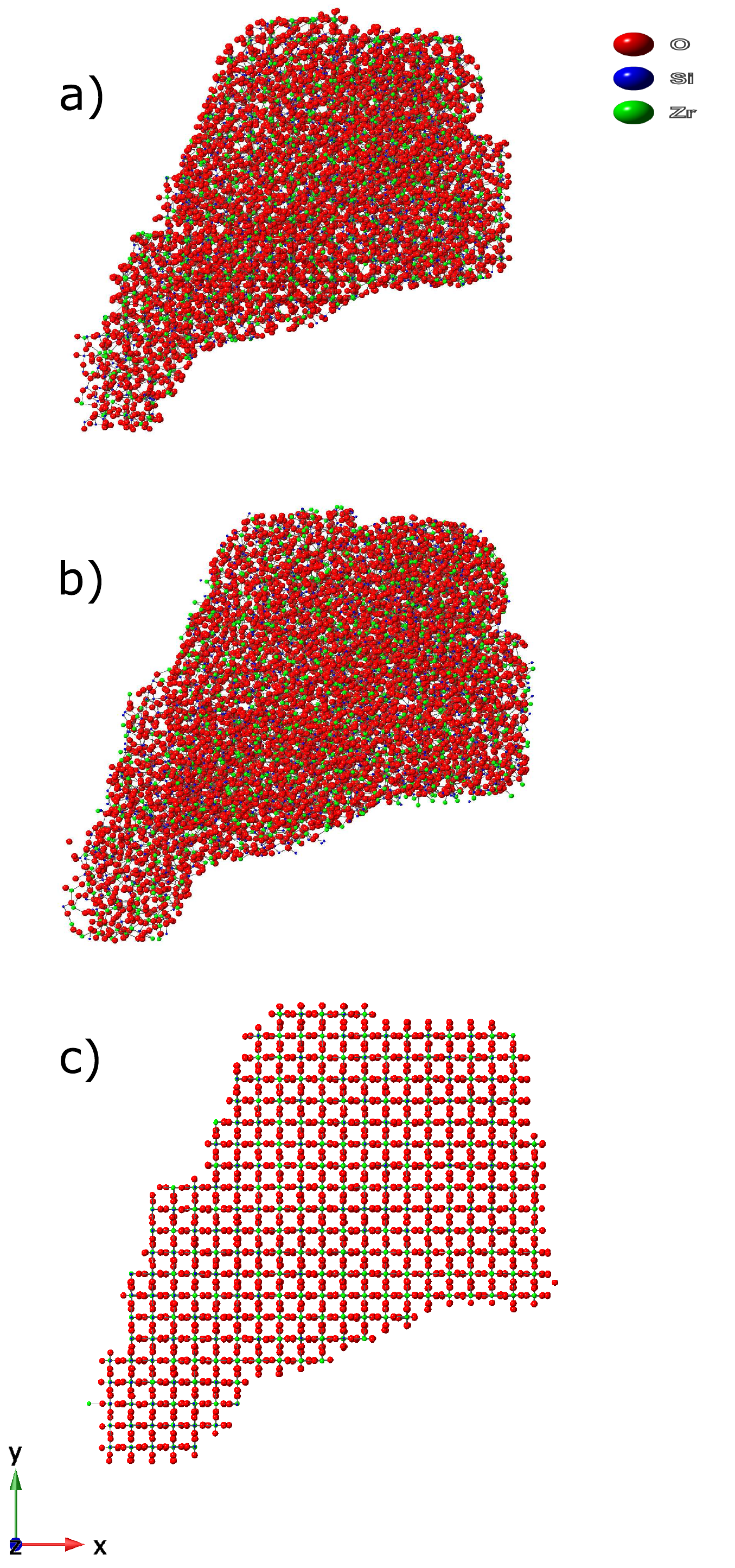}
\caption{ The atomic structures of a) the amorphised radiation damaged region in the pure crystalline zircon, b) a region in the melt-quench amorphous structure that is the same shape as the damaged region, c) a region in the crystalline zircon also of the same shape. These regions are used to calculate and fairly compare He diffusion between the various systems.}
\label{fig:damregion}
\end{figure}

\begin{figure}
\centering
\includegraphics[width = 12cm]{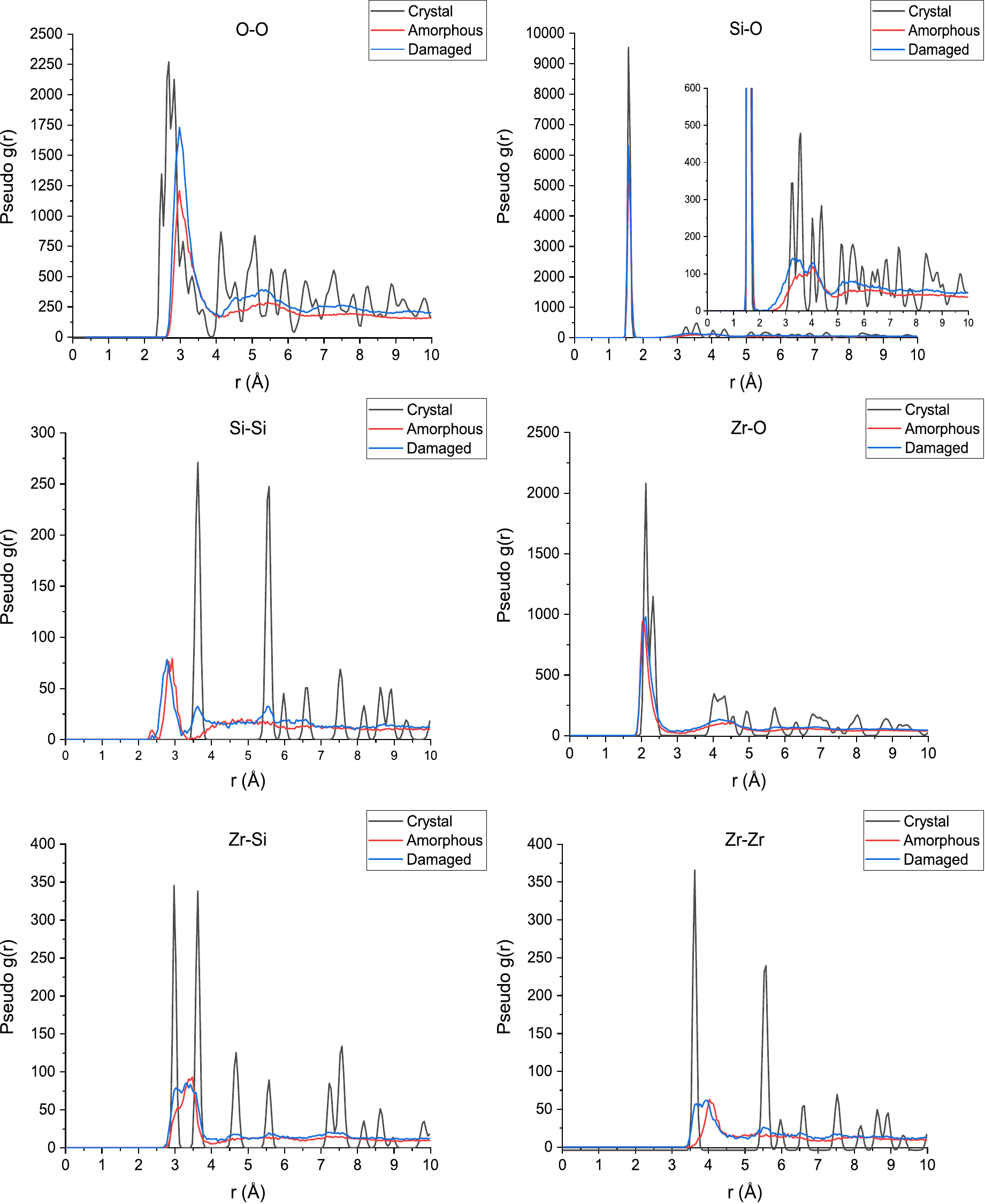}
\caption{Pseudo-partial pair distribution functions of the regions shown in Fig. \ref{fig:damregion}}
\label{fig:partials}
\end{figure}

This work uses the radiation damage structures produced in our previous paper \cite{diver2020evolution}. All the systems modelled contain 10 million atoms. There are 3 systems used: a pure crystal structure, a melt-quench amorphous structure produced using a cooling rate of 10 K/ps \cite{diver2020evolution}, and crystalline structure with radiation-damaged regions. The radiation damaged structure is produced as a result of 6 overlapping radiation cascades, each one due to a U ion with 70 keV of kinetic energy with trajectories along the same initial direction from the same initial position. The irregular shape of the  damaged region can be seen in Fig. \ref{fig:damregion}, with an estimated volume of 100,000 \r{A} \si{^3} . The damaged region was determined  empirically as a parallelepiped to include the full damaged region, such that there is a clear boundary visible between the crystal and the damaged region. The lack of long ranged order in the partial pair distribution functions (see Fig. \ref{fig:partials}) act as a further check that this region is sufficiently damaged compared to the rest of the structure.

These previously produced structures were then doped with helium atoms amounting to 1\% of the total atomic number of the original system. These were randomly placed in the existing simulation boxes. This allowed 100,000 He atoms to be sampled when calculating He diffusion in the amorphous and crystalline zircon structures. 

In order to calculate the diffusion separately in both the damaged and undamaged regions, and compare fairly between systems, the mean squared displacements (MSDs) of a subsample of the He atoms within the volume of the damaged region (approximately 250-400 at a time depending on temperature) were calculated. The He atoms only contributed to the MSD of the damaged region whilst within the volume, with its initial position within the region being recorded and these coordinates being where the MSD is calculated from. If the He atom were to leave and re-enter the system it would be considered as a new atom entering the subregion, with a new initial position. This means that any He paths outside the damaged region are not included in the calculation of the diffusion. The diffusion coefficients $D$ plotted are calculated during the diffusive regime of the helium atoms where the square of atomic displacement is linear with time. 

The Arrhenius equation  (Eq. \eqref{1}) relates diffusion to temperature and allows calculation of the activation energy for diffusion. 

\begin{equation}
    D = D_{0}e^{-\frac{E_a}{k_{\rm B}T}}
\label{1}
\end{equation}

where $D$ is the diffusion constant, $D_{0}$ the pre-exponential factor, $E_{a}$ is the activation energy required for diffusion and $T$ is the temperature. One can take the natural logarithm to yield 

\begin{equation}
    \ln (D)= \ln (D_{0}) -\frac{E_a}{k_{\rm B}T}
\label{2}
\end{equation}

\noindent and calculate $E_a$ from the slope of $\ln(D)$ vs $1/T$.

A series of MD simulations (8 in the crystal, 6 in the damaged system and 5 in the amorphous system) were run in each of the three systems at a range of temperatures from 1000 K to a maximum temperature of 2400 K in order to calculate the diffusion coefficients and the activation energy required for helium diffusion. For each temperature the systems were initially equilibrated for 100 ps followed by a 100 ps (80 ps for the damaged structure) run to collect statistics using the NPT hoover ensemble.

\section{Results and analysis}

\subsection{Diffusion in zircon}

\begin{figure}
\centering
\includegraphics[width = 12cm]{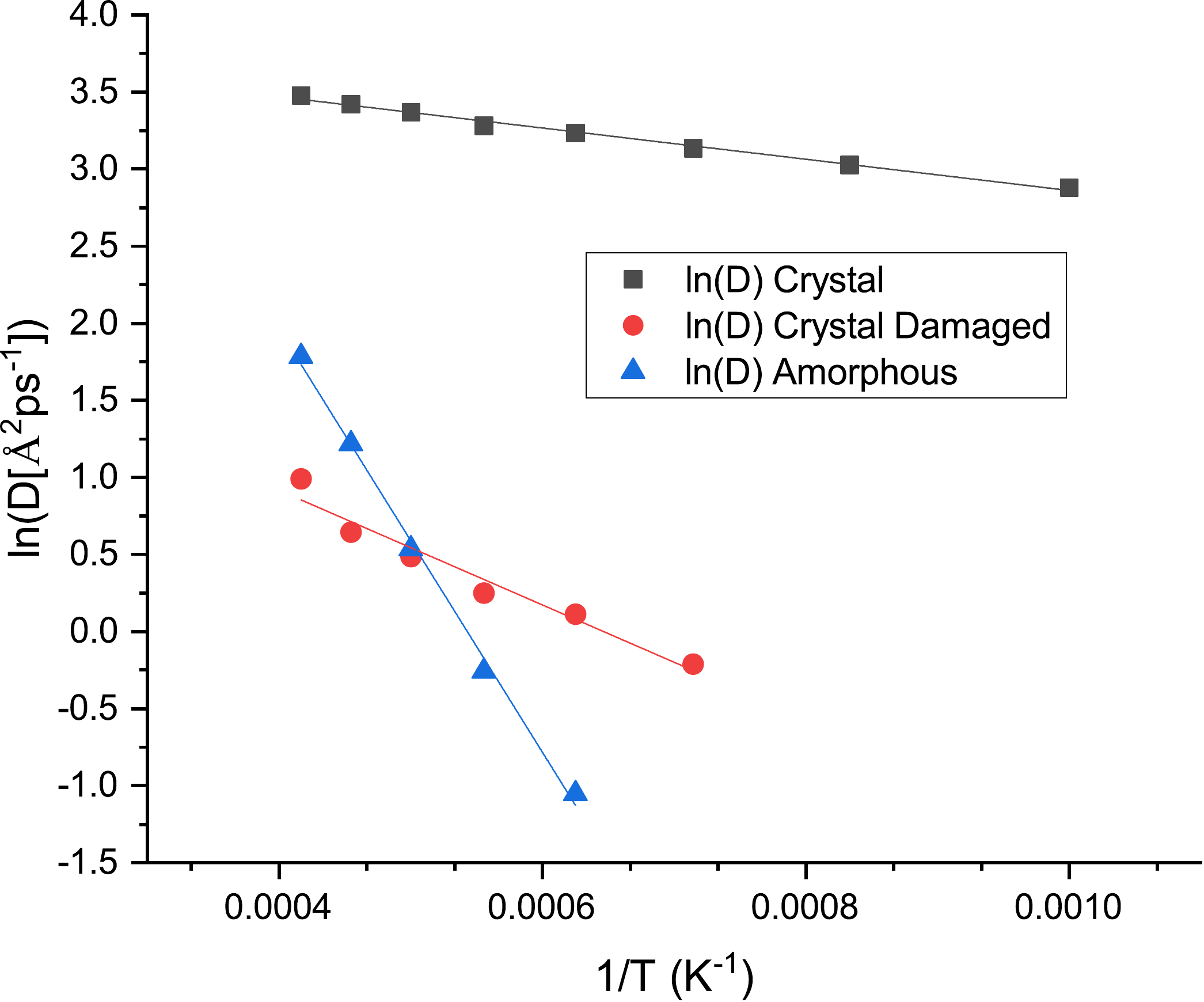}
\caption{Arrhenius plot for crystalline zircon, amorphous zircon and radiation damaged crystalline zircon}
\label{fig:arrplots}
\end{figure}

The Arrhenius plots produced for pure crystal, amorphous and radiation damaged zircon can be seen in Fig. \ref{fig:arrplots}. From the gradient of the relevant slopes in the Arrhenius plots a value for the activation energy of helium diffusion for the pure crystal, amorphous and damaged zircon structures are calculated (see Table \ref{tab:act_table} ). The $R^2$-values quantifying the fit of the Arrhenius plots to the straight line are in the range 0.96-0.99. The slight scatter seen in the damaged region is due to the smaller number of helium atoms sampled (250-400) in the damaged structure as compared to the 100,000 helium atoms sampled for the crystalline and amorphous simulations, because the radiation-damaged region is small compared to the total system size. 

\begin{table}
   \centering
    \begin{tabular}{|c|c|}
    \hline
      Structure  & Activation energy (kJ/mol)\\ \hline
      crystalline & 8.44 $\pm$ 0.26 \\ 
      amorphous  & 114 $\pm$ 3.22  \\ 
      damaged &  30.87 $\pm$ 3.93 \\ \hline
    \end{tabular}
\caption{Activation energies, calculated using MD, for He diffusion in crystalline, amorphous and damaged zircon.}
\label{tab:act_table}
\end{table}

As expected the crystal has the smallest activation energy due to the crystallographic pathways being both intact and not blocked at any points, allowing easy diffusion of the helium atoms. The activation energy of He diffusion in the damaged region of the crystalline structure is greater than that in the pure crystal, but is around a fourth of the activation energy required for He diffusion in the completely amorphous structure.


In our previous work \cite{diver2020evolution} we showed that the pair distribution function (PDF) of the most highly overlapped damaged region defined by a regular shape of a box containing 6,000 atoms was very close to the PDF of the fully amorphous structure. In order to study He diffusion in the damaged structure, we have used a larger region damaged by overlapping collision cascades. This region has an irregular shape as shown in Fig. 1a. To make a consistent comparison of the structural correlations in different regions measured by a distribution function, we have selected atoms in the melt-quench and crystalline structure bound by the same surface as in the radiation-damaged region (see Fig. 1b and 1c). We calculated the correlation function as $n(r)/r^2$, where $n(r)$ is the number of atoms in a bin, and refer to this function as ``pseudo-PDF'' because it is calculated for a finite non-periodic collection of atoms where the usual normalisation does not apply. 

The pseudo PDFs for all three of those regions are shown in Fig. \ref{fig:partials}. We observe the decrease and widening of PDF peaks in the disordered structures. We also observe a difference between the radiation-damaged structure and the fully amorphous one. For example, the radiation-structure shows fairly small peaks for O-O, Si-Si, Zr-Si and Zr-Zr sublattices in the medium-range order (e.g. second and third peaks) where no discernible peaks are seen in the fully amorphous structure. This is in agreement with our previous work \cite{diver2020evolution} in which it was shown that even after 6 collision cascades further structural damage and amorphization was still possible. As well as continual change to the local PDFs a continued increase in the number of coordination defects was also observed in the overlapped damaged region. The persistence of PDF peaks at the cations' crystalline peak positions have been also observed in amorphous zirconolite \cite{yang2014solid}. We will return to this point below when we discuss the variation of activation energies with the degree of introduced disorder. 

Our values for the activation energies in crystalline and amorphous zircon (see Table \ref{tab:act_table}) are the first we know of from large scale molecular dynamics simulations calculating bulk diffusion from a range of temperatures. As seen in Table \ref{tab:act_table} the calculated activation energy in the perfect crystal is 8.44 $\pm$ 0.26 kJ/mol. This is lower than experimentally measured values which are 10-20 times greater \cite{guenthner2013helium,cherniak2009diffusion,reiners2004zircon,reiners2005zircon}. However, the activation energy barriers for He diffusion calculated using a variety of computational methods, including DFT and molecular dynamics, are significantly lower than those measured in experiments. The modelling results also show a large variation of activation energy depending on the direction. For example, DFT nudged elastic band calculations of the activation energy barrier along the \textit{\textbf{c}} channel in crystalline zircon have reported values of 21.3 kJ/mol \cite{saadoune2009} and 42 kJ/mol \cite{bengtson2012he}. Similar calculations using DFT parameterized empirical potentials calculate the energy activation barrier along [001] to be 13.4 kJ/mol \cite{reich2007low}. Saadoune et al. \cite{saadoune2009} demonstrate that a single oxygen vacancy can lower the energy barrier for a jump from 21.3 kJ/mol to as little as 5.9 kJ/mol. The caveat is this subsequently acts as a trap for He, increasing the energy required for the next jump to 30 kJ/mol. Although the exact values for the He activation energy differ, most likely due to use of different functionals, step sizes and relaxation parameters for the surrounding lattice,  they all calculate the activation energy barrier perpendicular to the c-direction (i.e. along [100]) to be much higher, 255\cite{bengtson2012he}-259\cite{saadoune2009} kJ/mol. In an atomistic simulation study using pair wise potentials instead of DFT Saadoune et al. \cite{saadoune2009computer} showed the fastest diffusion pathway [100] to be along tetrahedral interstial sites in the perfect zircon lattice. This corresponds to an energy barrier for helium diffusion as low as 5 kJ/mol. This agrees well with our results, that in crystalline zircon the activation energy is low as diffusion is dominated by He pathways along the \textit{\textbf{c}} channel, and therefore increases as this channel is destroyed by radiation damage and subsequent complete amorphization.

Other experiments similarly report larger activation energies parallel to the c-channel as compared to simulations \cite{reiners2004zircon,reiners2005zircon}. In one study \cite{guenthner2013helium} where 7 samples are measured, activation energies parallel to the c-direction (in samples that are still crystalline) are reported to be between 138.22 kJ/mol and 166.19 kJ/mol, and those perpendicular to \textit{\textbf{c}} range from 106.52-169.75 kJ/mol. Two of the samples studied have no clear \textit{\textbf{c}} direction and are considered amorphized due to long term radiation damage and have activation energies of 145.96 and 70.76 kJ/mol.
In another experimental study \cite{cherniak2009diffusion} where He is implanted into polished zircon crystals via ion beam, the measured activation energies along each direction are almost equal, 148 kJ/mol parallel to the c-direction and 146 kJ/mol perpendicular. This discrepancy between experiment and the large number of different simulations carried out using various methodologies and techniques can be attributed to the large effects that even a small number of impurities, defects and radiation damage has on the diffusion of He along the c-channel. In the simulation studies zircon is a perfect crystal, which is impossible in the natural samples used experimentally where the zircon will contain Pb, Th and U impurities and will have experienced some level of radiation damage \cite{guenthner2013helium}. The use of ion beams to implant He into zircon also adds the possibility of atomic displacements and defects in the structure \cite{cherniak2009diffusion}. Saadoune et al. \cite{saadoune2009} show that the introduction of even a single oxygen vacancy along the channel can form a defect trap for He requiring over 30 keV to escape. Recent experimental work using laser depth profiling showed that the presence of said trace elements can have a significant effect on helium diffusion\cite{anderson2020helium}. An interesting experimental observation is that as the dose rate of natural samples goes from \num{1e16} $\alpha$/g to \num{1e19} $\alpha$/g, the resulting activation energy of helium diffusion falls from 166.19 kJ/mol to 70.7 kJ/mol due to the effect of damage percolation \cite{guenthner2013helium}. This is due to zircon being rendered totally amorphous at between \num{1e18} and \num{1e19} $\alpha$/g \cite{weber1998radiation}. The value calculated from our MD simulations for the fully amorphous structure (114 kJ/mol) is found to be between the activation energies of the two amorphous samples,\cite{guenthner2013helium} 145.96 kJ/mol and 70.7 kJ/mol respectively, which is in the range expected once zircon has been rendered completely amorphous by radiation damage.

The increase in activation energy due to small amounts of radiation damage or defects are corroborated by the only (to the authors' knowledge) computational work that models the effect of damage on diffusion in zircon \cite{gautheron2020multi}. Gautheron et al. \cite{gautheron2020multi} use a kinetic Monte-Carlo (KMC) model based on DFT calculations of activation energy and attempt rates where the helium is modelled jumping  between interstitial sites. The effect of ``damage'' is modelled by randomly blocking some percentage of the interstitial sites. Even after blocking only 1 percent of interstitial sites they calculate a large increase in the activation energy from 23 kJ/mol in the perfect crystal to around 60 kJ/mol, around the half value we calculated for the completely amorphous structure and within 4 kJ/mol of the activation energy along the perpendicular directions to c. Additional blocking of sites, up to 20\%, does not increase the activation energy calculated using KMC. Whilst we calculated the activation energy in the damaged region after 6 radiation cascades to be 30.87 $\pm 3.93$ kJ/mol, half that of the value Gatheron et al. \cite{gautheron2020multi} calculated for 'damaged' zircon, consistently we also see a three-fold increased in the value of activation energy in the damaged region compared to the crystal. The lower estimate of the activation energy calculated using MD compared to DFT may include different factors. For example, DFT does not account for temperature and collective modes facilitating diffusion. Another difference between this and previously mentioned modelling studies on one hand and our current simulations on the other hand is that we do not specify a certain direction or pathway of diffusion: our activation energies should be viewed as effective values with contributions from all possible pathways. Nevertheless, the same trends are observed in both studies, that radiation damage significantly increases the activation energy in zircon, even before the complete loss of the cation sublattice.

In the above discussion, decreasing crystallinity results in the disappearance of fast diffusion pathways. This is consistent with (a) our finding that the activation energy is larger in the fully amorphous structure as compared to radiation-damaged one and (b) our earlier observation that the radiation-damaged structure has more order than the fully amorphous systems: recall that the radiation-structure shows fairly small peaks for O-O, Si-Si, Zr-Si and Zr-Zr sublattices in the medium-range order (e.g. second and third peaks) where no discernable peaks are seen in the fully amorphous structure.

To summarise, the main result of this section is the increase of the activation energy with the amount of introduced disorder.

\subsection{Helium accumulation in damaged zircon}

\begin{figure}
\centering
\includegraphics[width = 8cm]{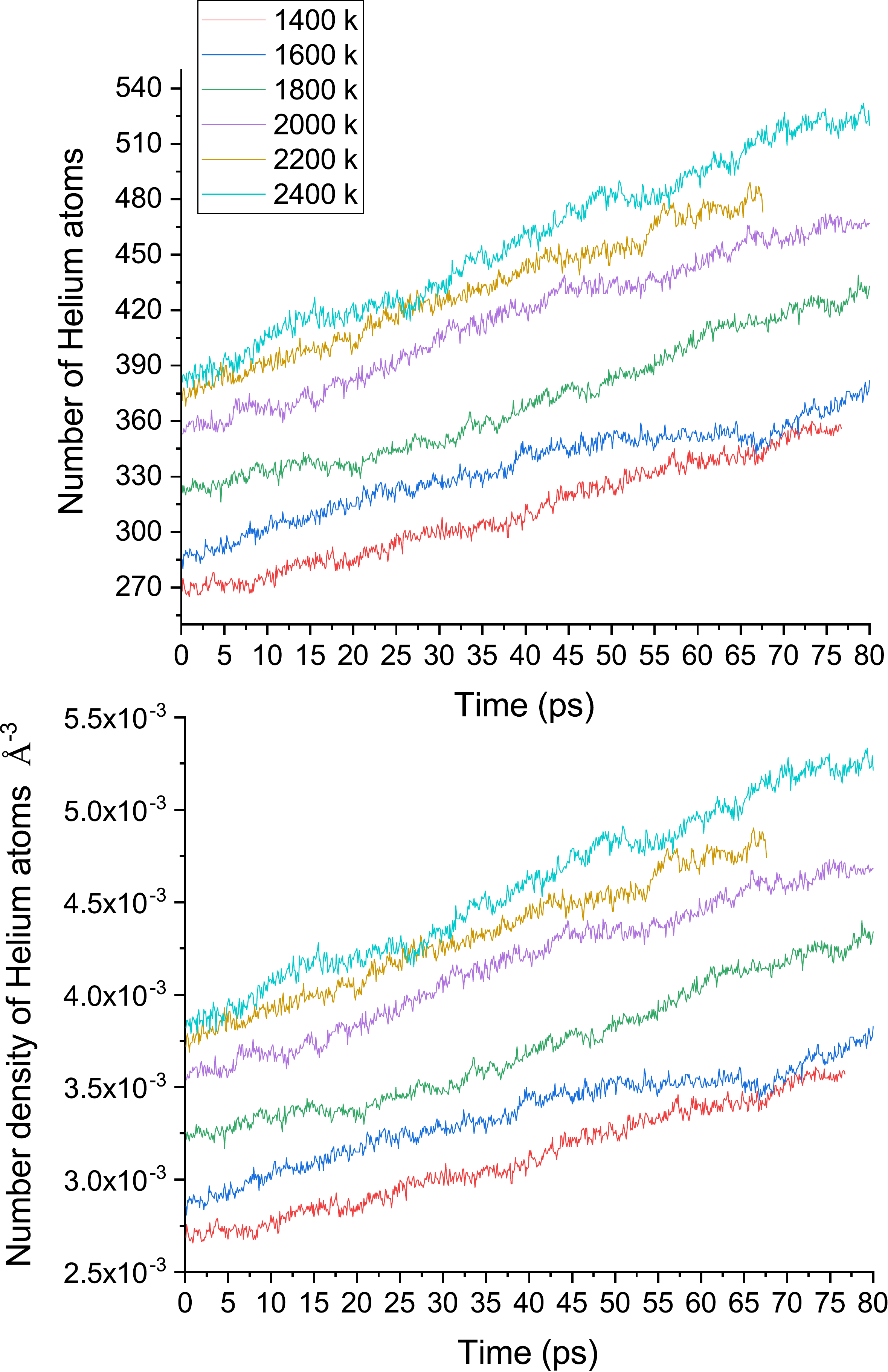}
\caption{The number of helium atoms and the number density of helium atoms inside the damaged region against time for a series of temperatures. The initial high value is due to accumulation during the equilibration phase. }
\label{fig:heliumacum}
\end{figure}

Another effect of radiation damage to consider is the accumulation of helium inside the damaged region. The diffusion rate in the crystalline region is much higher, and as such He will arrive at the damaged region much faster than it will exit, leading to a build up. Fig. \ref{fig:heliumacum} shows the number of helium atoms inside the damaged region during the temperature runs, as well as the number density of He in the damaged region, against time. The He number density in the damaged region at each temperature can be compared to the average number density of the whole system at each temperature shown in Table \ref{tab:He_tablel}, where the number density ranges from 0.825-0.848$\times10^{-3}$\r{A}$^{-3}$ in the crystal and 0.694-0.743$\times10^{-3}$\r{A}$^{-3}$ in amorphous zircon. When comparing the number densities of helium atoms in the damaged region to the whole system it is evident there is significantly more helium in that region in Fig. \ref{fig:heliumacum} than in both the amorphous and crystalline structures in Table \ref{tab:He_tablel}, and that the number density continues to increase with time. We note that we do not observe saturation of the number of He atoms in the damaged region due to limitations related to simulating large system sizes for long times.

Our finding indicates that during the process of damage and amorphization helium atoms will be mainly trapped in the damaged region and likely remain trapped there for longer due to increased activation energies. This raises the possibility that He will become inhomogeneously stored in zircon, becoming concentrated in regions that are damaged first.

\begin{table}
   \centering
    \begin{tabular}{|c|c|c|}
    \hline
      Temperature (k) &  {Number density of helium atoms (\r{A}$^{-3}$) crystal} & {Number density of helium atoms (\r{A}$^{-3}$) Amorphous} \\ \hline
     
    1400 & \num{8.476e-4} & \num{7.433e-4}\\ \hline
    1600 & \num{8.434e-4} & \num{7.387e-4}\\ \hline
    1800 & \num{8.391e-4} & \num{7.315e-4}\\ \hline
    2000 & \num{8.346e-4} & \num{7.216e-4}\\ \hline
    2200 & \num{8.314e-4} & \num{7.091e-4}\\ \hline
    2400 & \num{8.246e-4} & \num{6.944e-4}\\ \hline
    
    \end{tabular}
    \caption{Number density of helium atoms in amorphous and crystalline zircon}
    \label{tab:He_tablel}
\end{table}

Although from irradiation experiments on zircon no helium bubble formation was reported from electron microscope analysis \cite{weber1993alpha}, it is known that the formation of helium bubbles occurs in other materials studied as potential waste forms. This is particularly apparent in euxenite, were helium bubble formation occurs in high enough concentrations that it can be related to swelling and eventually cracking \cite{seydoux2016trapping}. Recent work on zirconolite showed that helium bubble formation occurs in the amorphous damaged phase and leaching would be expected to occur in a helium bubble containing zirconolite waste form \cite{mir2021situ}. Our simulations suggest that He build up occurs in radiation-damaged parts of a waste form, a prediction that can be studied experimentally.

\section{Conclusions}
 
Using molecular dynamic simulations, we have modelled the diffusion of helium atoms in crystalline and amorphous zircon and radiation damaged crystalline zircon. Examination of the PDFs of the radiation damaged zircon structure shows that there is some retained order in the cation sublattice. This retained 'ordering' plays a large part in the diffusion of helium, leading to a lower activation energy in a radiation damaged zircon region compared to that of a fully amorphous structure generated using the melt-quench method. 

We have also for the first time reported activation energys for helium diffusion using molecular dynamics taking into account a 1\% doping of helium atoms in a crystal, damaged and amorphous zircon structures, which is consistent with results from different computational methods used previously. We also show good agreement with experimentally measured activation energies in amorphous zircon\cite{gautheron2020multi}. 

The large difference in diffusion rates and activation energies calculated in crystalline and damaged zircon is further evidenced by the build up of He that we predict in the damaged region via our MD simulations. Understanding how this build up may effect bubble formation or cracking in zircon is of importance for predicting waste form performance in long term.

\section{Acknowledgements}

We are grateful to A. Mir for discussions. 

AD, OD and KT were supported by the UK Engineering and Physical Sciences Research Council (EPSRC) grant EP/R004870/1.

Via our membership of the UK's HEC Materials Chemistry Consortium, which is funded by EPSRC (EP/L000202, EP/R029431), this work used the ARCHER UK National Supercomputing Service (http://www.archer.ac.uk) and the UK Materials and Molecular Modelling Hub for computational resources, MMM Hub, which is partially funded by EPSRC (EP/P020194). Part of this work made use of computational support by CoSeC, the Computational Science Centre for Research Communities, through CCP5: The Computer Simulation of Condensed Phases, EPSRC grant no EP/M022617/1 through AME and ITT.

\bibliographystyle{apsrev4-2}
\bibliography{He}

\begin{thebibliography}{33}%
\makeatletter
\providecommand \@ifxundefined [1]{%
 \@ifx{#1\undefined}
}%
\providecommand \@ifnum [1]{%
 \ifnum #1\expandafter \@firstoftwo
 \else \expandafter \@secondoftwo
 \fi
}%
\providecommand \@ifx [1]{%
 \ifx #1\expandafter \@firstoftwo
 \else \expandafter \@secondoftwo
 \fi
}%
\providecommand \natexlab [1]{#1}%
\providecommand \enquote  [1]{``#1''}%
\providecommand \bibnamefont  [1]{#1}%
\providecommand \bibfnamefont [1]{#1}%
\providecommand \citenamefont [1]{#1}%
\providecommand \href@noop [0]{\@secondoftwo}%
\providecommand \href [0]{\begingroup \@sanitize@url \@href}%
\providecommand \@href[1]{\@@startlink{#1}\@@href}%
\providecommand \@@href[1]{\endgroup#1\@@endlink}%
\providecommand \@sanitize@url [0]{\catcode `\\12\catcode `\$12\catcode
  `\&12\catcode `\#12\catcode `\^12\catcode `\_12\catcode `\%12\relax}%
\providecommand \@@startlink[1]{}%
\providecommand \@@endlink[0]{}%
\providecommand \url  [0]{\begingroup\@sanitize@url \@url }%
\providecommand \@url [1]{\endgroup\@href {#1}{\urlprefix }}%
\providecommand \urlprefix  [0]{URL }%
\providecommand \Eprint [0]{\href }%
\providecommand \doibase [0]{https://doi.org/}%
\providecommand \selectlanguage [0]{\@gobble}%
\providecommand \bibinfo  [0]{\@secondoftwo}%
\providecommand \bibfield  [0]{\@secondoftwo}%
\providecommand \translation [1]{[#1]}%
\providecommand \BibitemOpen [0]{}%
\providecommand \bibitemStop [0]{}%
\providecommand \bibitemNoStop [0]{.\EOS\space}%
\providecommand \EOS [0]{\spacefactor3000\relax}%
\providecommand \BibitemShut  [1]{\csname bibitem#1\endcsname}%
\let\auto@bib@innerbib\@empty
\bibitem [{\citenamefont {Ewing}\ \emph {et~al.}(1995)\citenamefont {Ewing},
  \citenamefont {Lutze},\ and\ \citenamefont {Weber}}]{ewing_lutze_weber_1995}%
  \BibitemOpen
  \bibfield  {author} {\bibinfo {author} {\bibfnamefont {R.}~\bibnamefont
  {Ewing}}, \bibinfo {author} {\bibfnamefont {W.}~\bibnamefont {Lutze}},\ and\
  \bibinfo {author} {\bibfnamefont {W.~J.}\ \bibnamefont {Weber}},\ }\href
  {https://doi.org/10.1557/JMR.1995.0243} {\bibfield  {journal} {\bibinfo
  {journal} {Journal of Materials Research}\ }\textbf {\bibinfo {volume}
  {10}},\ \bibinfo {pages} {243–246} (\bibinfo {year} {1995})}\BibitemShut
  {NoStop}%
\bibitem [{\citenamefont {Ewing}\ and\ \citenamefont
  {Wang}(1994)}]{ewing1994radiation}%
  \BibitemOpen
  \bibfield  {author} {\bibinfo {author} {\bibfnamefont {R.~C.}\ \bibnamefont
  {Ewing}}\ and\ \bibinfo {author} {\bibfnamefont {L.-M.}\ \bibnamefont
  {Wang}},\ }\href@noop {} {\bibfield  {journal} {\bibinfo  {journal} {J.
  Mater. res}\ }\textbf {\bibinfo {volume} {9}},\ \bibinfo {pages} {889}
  (\bibinfo {year} {1994})}\BibitemShut {NoStop}%
\bibitem [{\citenamefont {Murakami}\ \emph {et~al.}(1991)\citenamefont
  {Murakami}, \citenamefont {Chakoumakos}, \citenamefont {Ewing}, \citenamefont
  {Lumpkin},\ and\ \citenamefont {Weber}}]{murakami1991alpha}%
  \BibitemOpen
  \bibfield  {author} {\bibinfo {author} {\bibfnamefont {T.}~\bibnamefont
  {Murakami}}, \bibinfo {author} {\bibfnamefont {B.~C.}\ \bibnamefont
  {Chakoumakos}}, \bibinfo {author} {\bibfnamefont {R.~C.}\ \bibnamefont
  {Ewing}}, \bibinfo {author} {\bibfnamefont {G.~R.}\ \bibnamefont {Lumpkin}},\
  and\ \bibinfo {author} {\bibfnamefont {W.~J.}\ \bibnamefont {Weber}},\
  }\href@noop {} {\bibfield  {journal} {\bibinfo  {journal} {American
  Mineralogist}\ }\textbf {\bibinfo {volume} {76}},\ \bibinfo {pages} {1510}
  (\bibinfo {year} {1991})}\BibitemShut {NoStop}%
\bibitem [{\citenamefont {Weber}(1990)}]{weber1990radiation}%
  \BibitemOpen
  \bibfield  {author} {\bibinfo {author} {\bibfnamefont {W.}~\bibnamefont
  {Weber}},\ }\href@noop {} {\bibfield  {journal} {\bibinfo  {journal} {Journal
  of Materials Research}\ }\textbf {\bibinfo {volume} {5}},\ \bibinfo {pages}
  {2687} (\bibinfo {year} {1990})}\BibitemShut {NoStop}%
\bibitem [{\citenamefont {Weber}(1991)}]{weber1991self}%
  \BibitemOpen
  \bibfield  {author} {\bibinfo {author} {\bibfnamefont {W.}~\bibnamefont
  {Weber}},\ }\href@noop {} {\bibfield  {journal} {\bibinfo  {journal}
  {Radiation Effects and Defects in Solids}\ }\textbf {\bibinfo {volume}
  {115}},\ \bibinfo {pages} {341} (\bibinfo {year} {1991})}\BibitemShut
  {NoStop}%
\bibitem [{\citenamefont {Weber}\ \emph {et~al.}(1994)\citenamefont {Weber},
  \citenamefont {Ewing},\ and\ \citenamefont {Wang}}]{weber1994radiation}%
  \BibitemOpen
  \bibfield  {author} {\bibinfo {author} {\bibfnamefont {W.~J.}\ \bibnamefont
  {Weber}}, \bibinfo {author} {\bibfnamefont {R.~C.}\ \bibnamefont {Ewing}},\
  and\ \bibinfo {author} {\bibfnamefont {L.-M.}\ \bibnamefont {Wang}},\
  }\href@noop {} {\bibfield  {journal} {\bibinfo  {journal} {Journal of
  Materials Research}\ }\textbf {\bibinfo {volume} {9}},\ \bibinfo {pages}
  {688} (\bibinfo {year} {1994})}\BibitemShut {NoStop}%
\bibitem [{\citenamefont {Trachenko}\ \emph {et~al.}(2004)\citenamefont
  {Trachenko}, \citenamefont {Dove}, \citenamefont {Geisler}, \citenamefont
  {Todorov},\ and\ \citenamefont {Smith}}]{Trachenko_2004per}%
  \BibitemOpen
  \bibfield  {author} {\bibinfo {author} {\bibfnamefont {K.}~\bibnamefont
  {Trachenko}}, \bibinfo {author} {\bibfnamefont {M.~T.}\ \bibnamefont {Dove}},
  \bibinfo {author} {\bibfnamefont {T.}~\bibnamefont {Geisler}}, \bibinfo
  {author} {\bibfnamefont {I.}~\bibnamefont {Todorov}},\ and\ \bibinfo {author}
  {\bibfnamefont {B.}~\bibnamefont {Smith}},\ }\href
  {https://doi.org/10.1088/0953-8984/16/27/002} {\bibfield  {journal} {\bibinfo
   {journal} {Journal of Physics: Condensed Matter}\ }\textbf {\bibinfo
  {volume} {16}},\ \bibinfo {pages} {S2623} (\bibinfo {year}
  {2004})}\BibitemShut {NoStop}%
\bibitem [{\citenamefont {Trachenko}\ \emph {et~al.}(2002)\citenamefont
  {Trachenko}, \citenamefont {Dove},\ and\ \citenamefont
  {Salje}}]{PhysRevB.65.180102}%
  \BibitemOpen
  \bibfield  {author} {\bibinfo {author} {\bibfnamefont {K.}~\bibnamefont
  {Trachenko}}, \bibinfo {author} {\bibfnamefont {M.~T.}\ \bibnamefont
  {Dove}},\ and\ \bibinfo {author} {\bibfnamefont {E.~K.~H.}\ \bibnamefont
  {Salje}},\ }\href {https://doi.org/10.1103/PhysRevB.65.180102} {\bibfield
  {journal} {\bibinfo  {journal} {Phys. Rev. B}\ }\textbf {\bibinfo {volume}
  {65}},\ \bibinfo {pages} {180102} (\bibinfo {year} {2002})}\BibitemShut
  {NoStop}%
\bibitem [{\citenamefont {Diver}\ \emph {et~al.}(2020)\citenamefont {Diver},
  \citenamefont {Dicks}, \citenamefont {Trachenko}, \citenamefont {Elena},\
  and\ \citenamefont {Todorov}}]{diver2020evolution}%
  \BibitemOpen
  \bibfield  {author} {\bibinfo {author} {\bibfnamefont {A.}~\bibnamefont
  {Diver}}, \bibinfo {author} {\bibfnamefont {O.}~\bibnamefont {Dicks}},
  \bibinfo {author} {\bibfnamefont {K.}~\bibnamefont {Trachenko}}, \bibinfo
  {author} {\bibfnamefont {A.}~\bibnamefont {Elena}},\ and\ \bibinfo {author}
  {\bibfnamefont {I.~T.}\ \bibnamefont {Todorov}},\ }\href
  {http://iopscience.iop.org/10.1088/1361-648X/ab9f51} {\bibfield  {journal}
  {\bibinfo  {journal} {Journal of Physics: Condensed Matter}\ } (\bibinfo
  {year} {2020})}\BibitemShut {NoStop}%
\bibitem [{\citenamefont {Weber}\ and\ \citenamefont
  {Roberts}(1983)}]{weber1983review}%
  \BibitemOpen
  \bibfield  {author} {\bibinfo {author} {\bibfnamefont {W.~J.}\ \bibnamefont
  {Weber}}\ and\ \bibinfo {author} {\bibfnamefont {F.~P.}\ \bibnamefont
  {Roberts}},\ }\href@noop {} {\bibfield  {journal} {\bibinfo  {journal}
  {Nuclear Technology}\ }\textbf {\bibinfo {volume} {60}},\ \bibinfo {pages}
  {178} (\bibinfo {year} {1983})}\BibitemShut {NoStop}%
\bibitem [{\citenamefont {Weber}\ \emph {et~al.}(1998)\citenamefont {Weber},
  \citenamefont {Ewing}, \citenamefont {Catlow}, \citenamefont {De~La~Rubia},
  \citenamefont {Hobbs}, \citenamefont {Kinoshita}, \citenamefont {Motta},
  \citenamefont {Nastasi}, \citenamefont {Salje}, \citenamefont {Vance} \emph
  {et~al.}}]{weber1998radiation}%
  \BibitemOpen
  \bibfield  {author} {\bibinfo {author} {\bibfnamefont {W.}~\bibnamefont
  {Weber}}, \bibinfo {author} {\bibfnamefont {R.}~\bibnamefont {Ewing}},
  \bibinfo {author} {\bibfnamefont {C.}~\bibnamefont {Catlow}}, \bibinfo
  {author} {\bibfnamefont {T.~D.}\ \bibnamefont {De~La~Rubia}}, \bibinfo
  {author} {\bibfnamefont {L.}~\bibnamefont {Hobbs}}, \bibinfo {author}
  {\bibfnamefont {C.}~\bibnamefont {Kinoshita}}, \bibinfo {author}
  {\bibfnamefont {A.}~\bibnamefont {Motta}}, \bibinfo {author} {\bibfnamefont
  {M.}~\bibnamefont {Nastasi}}, \bibinfo {author} {\bibfnamefont
  {E.}~\bibnamefont {Salje}}, \bibinfo {author} {\bibfnamefont
  {E.}~\bibnamefont {Vance}}, \emph {et~al.},\ }\href@noop {} {\bibfield
  {journal} {\bibinfo  {journal} {Journal of Materials Research}\ }\textbf
  {\bibinfo {volume} {13}},\ \bibinfo {pages} {1434} (\bibinfo {year}
  {1998})}\BibitemShut {NoStop}%
\bibitem [{\citenamefont {Headley}\ \emph {et~al.}(1981)\citenamefont
  {Headley}, \citenamefont {Ewing},\ and\ \citenamefont
  {Haaker}}]{headley1981amorphous}%
  \BibitemOpen
  \bibfield  {author} {\bibinfo {author} {\bibfnamefont {T.}~\bibnamefont
  {Headley}}, \bibinfo {author} {\bibfnamefont {R.~C.}\ \bibnamefont {Ewing}},\
  and\ \bibinfo {author} {\bibfnamefont {R.~F.}\ \bibnamefont {Haaker}},\
  }\href@noop {} {\bibfield  {journal} {\bibinfo  {journal} {Nature}\ }\textbf
  {\bibinfo {volume} {293}},\ \bibinfo {pages} {449} (\bibinfo {year}
  {1981})}\BibitemShut {NoStop}%
\bibitem [{\citenamefont {Lumpkin}\ \emph {et~al.}(1986)\citenamefont
  {Lumpkin}, \citenamefont {Ewing}, \citenamefont {Chakoumakos}, \citenamefont
  {Greegor}, \citenamefont {Lytle}, \citenamefont {Foltyn}, \citenamefont
  {Clinard}, \citenamefont {Boatner},\ and\ \citenamefont
  {Abraham}}]{lumpkin1986alpha}%
  \BibitemOpen
  \bibfield  {author} {\bibinfo {author} {\bibfnamefont {G.}~\bibnamefont
  {Lumpkin}}, \bibinfo {author} {\bibfnamefont {R.}~\bibnamefont {Ewing}},
  \bibinfo {author} {\bibfnamefont {B.}~\bibnamefont {Chakoumakos}}, \bibinfo
  {author} {\bibfnamefont {R.}~\bibnamefont {Greegor}}, \bibinfo {author}
  {\bibfnamefont {F.}~\bibnamefont {Lytle}}, \bibinfo {author} {\bibfnamefont
  {E.}~\bibnamefont {Foltyn}}, \bibinfo {author} {\bibfnamefont
  {F.}~\bibnamefont {Clinard}}, \bibinfo {author} {\bibfnamefont
  {L.}~\bibnamefont {Boatner}},\ and\ \bibinfo {author} {\bibfnamefont
  {M.}~\bibnamefont {Abraham}},\ }\href@noop {} {\bibfield  {journal} {\bibinfo
   {journal} {Journal of Materials Research}\ }\textbf {\bibinfo {volume}
  {1}},\ \bibinfo {pages} {564} (\bibinfo {year} {1986})}\BibitemShut {NoStop}%
\bibitem [{\citenamefont {Seydoux-Guillaume}\ \emph {et~al.}(2016)\citenamefont
  {Seydoux-Guillaume}, \citenamefont {David}, \citenamefont {Alix},
  \citenamefont {Datas},\ and\ \citenamefont {Bingen}}]{seydoux2016trapping}%
  \BibitemOpen
  \bibfield  {author} {\bibinfo {author} {\bibfnamefont {A.-M.}\ \bibnamefont
  {Seydoux-Guillaume}}, \bibinfo {author} {\bibfnamefont {M.-L.}\ \bibnamefont
  {David}}, \bibinfo {author} {\bibfnamefont {K.}~\bibnamefont {Alix}},
  \bibinfo {author} {\bibfnamefont {L.}~\bibnamefont {Datas}},\ and\ \bibinfo
  {author} {\bibfnamefont {B.}~\bibnamefont {Bingen}},\ }\href@noop {}
  {\bibfield  {journal} {\bibinfo  {journal} {Earth and Planetary Science
  Letters}\ }\textbf {\bibinfo {volume} {448}},\ \bibinfo {pages} {133}
  (\bibinfo {year} {2016})}\BibitemShut {NoStop}%
\bibitem [{\citenamefont {Reiners}(2005)}]{reiners2005zircon}%
  \BibitemOpen
  \bibfield  {author} {\bibinfo {author} {\bibfnamefont {P.~W.}\ \bibnamefont
  {Reiners}},\ }\href@noop {} {\bibfield  {journal} {\bibinfo  {journal}
  {Reviews in Mineralogy and Geochemistry}\ }\textbf {\bibinfo {volume} {58}},\
  \bibinfo {pages} {151} (\bibinfo {year} {2005})}\BibitemShut {NoStop}%
\bibitem [{\citenamefont {Reiners}\ \emph {et~al.}(2004)\citenamefont
  {Reiners}, \citenamefont {Spell}, \citenamefont {Nicolescu},\ and\
  \citenamefont {Zanetti}}]{reiners2004zircon}%
  \BibitemOpen
  \bibfield  {author} {\bibinfo {author} {\bibfnamefont {P.~W.}\ \bibnamefont
  {Reiners}}, \bibinfo {author} {\bibfnamefont {T.~L.}\ \bibnamefont {Spell}},
  \bibinfo {author} {\bibfnamefont {S.}~\bibnamefont {Nicolescu}},\ and\
  \bibinfo {author} {\bibfnamefont {K.~A.}\ \bibnamefont {Zanetti}},\
  }\href@noop {} {\bibfield  {journal} {\bibinfo  {journal} {Geochimica et
  cosmochimica acta}\ }\textbf {\bibinfo {volume} {68}},\ \bibinfo {pages}
  {1857} (\bibinfo {year} {2004})}\BibitemShut {NoStop}%
\bibitem [{\citenamefont {Bengtson}\ \emph {et~al.}(2012)\citenamefont
  {Bengtson}, \citenamefont {Ewing},\ and\ \citenamefont
  {Becker}}]{bengtson2012he}%
  \BibitemOpen
  \bibfield  {author} {\bibinfo {author} {\bibfnamefont {A.}~\bibnamefont
  {Bengtson}}, \bibinfo {author} {\bibfnamefont {R.~C.}\ \bibnamefont
  {Ewing}},\ and\ \bibinfo {author} {\bibfnamefont {U.}~\bibnamefont
  {Becker}},\ }\href@noop {} {\bibfield  {journal} {\bibinfo  {journal}
  {Geochimica et Cosmochimica Acta}\ }\textbf {\bibinfo {volume} {86}},\
  \bibinfo {pages} {228} (\bibinfo {year} {2012})}\BibitemShut {NoStop}%
\bibitem [{\citenamefont {Farley}(2007)}]{farley2007he}%
  \BibitemOpen
  \bibfield  {author} {\bibinfo {author} {\bibfnamefont {K.}~\bibnamefont
  {Farley}},\ }\href@noop {} {\bibfield  {journal} {\bibinfo  {journal}
  {Geochimica et Cosmochimica Acta}\ }\textbf {\bibinfo {volume} {71}},\
  \bibinfo {pages} {4015} (\bibinfo {year} {2007})}\BibitemShut {NoStop}%
\bibitem [{\citenamefont {Guenthner}\ \emph {et~al.}(2013)\citenamefont
  {Guenthner}, \citenamefont {Reiners}, \citenamefont {Ketcham}, \citenamefont
  {Nasdala},\ and\ \citenamefont {Giester}}]{guenthner2013helium}%
  \BibitemOpen
  \bibfield  {author} {\bibinfo {author} {\bibfnamefont {W.~R.}\ \bibnamefont
  {Guenthner}}, \bibinfo {author} {\bibfnamefont {P.~W.}\ \bibnamefont
  {Reiners}}, \bibinfo {author} {\bibfnamefont {R.~A.}\ \bibnamefont
  {Ketcham}}, \bibinfo {author} {\bibfnamefont {L.}~\bibnamefont {Nasdala}},\
  and\ \bibinfo {author} {\bibfnamefont {G.}~\bibnamefont {Giester}},\
  }\href@noop {} {\bibfield  {journal} {\bibinfo  {journal} {American Journal
  of Science}\ }\textbf {\bibinfo {volume} {313}},\ \bibinfo {pages} {145}
  (\bibinfo {year} {2013})}\BibitemShut {NoStop}%
\bibitem [{\citenamefont {Cherniak}\ \emph {et~al.}(2009)\citenamefont
  {Cherniak}, \citenamefont {Watson},\ and\ \citenamefont
  {Thomas}}]{cherniak2009diffusion}%
  \BibitemOpen
  \bibfield  {author} {\bibinfo {author} {\bibfnamefont {D.~J.}\ \bibnamefont
  {Cherniak}}, \bibinfo {author} {\bibfnamefont {E.~B.}\ \bibnamefont
  {Watson}},\ and\ \bibinfo {author} {\bibfnamefont {J.~B.}\ \bibnamefont
  {Thomas}},\ }\href {https://doi.org/10.1016/j.chemgeo.2009.08.011} {\bibfield
   {journal} {\bibinfo  {journal} {Chemical Geology}\ }\textbf {\bibinfo
  {volume} {268}},\ \bibinfo {pages} {155} (\bibinfo {year}
  {2009})}\BibitemShut {NoStop}%
\bibitem [{\citenamefont {Dani{\v{s}}{\'\i}k}\ \emph
  {et~al.}(2017)\citenamefont {Dani{\v{s}}{\'\i}k}, \citenamefont {McInnes},
  \citenamefont {Kirkland}, \citenamefont {McDonald}, \citenamefont {Evans},\
  and\ \citenamefont {Becker}}]{danivsik2017seeing}%
  \BibitemOpen
  \bibfield  {author} {\bibinfo {author} {\bibfnamefont {M.}~\bibnamefont
  {Dani{\v{s}}{\'\i}k}}, \bibinfo {author} {\bibfnamefont {B.~I.}\ \bibnamefont
  {McInnes}}, \bibinfo {author} {\bibfnamefont {C.~L.}\ \bibnamefont
  {Kirkland}}, \bibinfo {author} {\bibfnamefont {B.~J.}\ \bibnamefont
  {McDonald}}, \bibinfo {author} {\bibfnamefont {N.~J.}\ \bibnamefont
  {Evans}},\ and\ \bibinfo {author} {\bibfnamefont {T.}~\bibnamefont
  {Becker}},\ }\href@noop {} {\bibfield  {journal} {\bibinfo  {journal}
  {Science advances}\ }\textbf {\bibinfo {volume} {3}},\ \bibinfo {pages}
  {e1601121} (\bibinfo {year} {2017})}\BibitemShut {NoStop}%
\bibitem [{\citenamefont {Gautheron}\ \emph {et~al.}(2020)\citenamefont
  {Gautheron}, \citenamefont {Djimbi}, \citenamefont {Roques}, \citenamefont
  {Balout}, \citenamefont {Ketcham}, \citenamefont {Simoni}, \citenamefont
  {Pik}, \citenamefont {Seydoux-Guillaume},\ and\ \citenamefont
  {Tassan-Got}}]{gautheron2020multi}%
  \BibitemOpen
  \bibfield  {author} {\bibinfo {author} {\bibfnamefont {C.}~\bibnamefont
  {Gautheron}}, \bibinfo {author} {\bibfnamefont {D.~M.}\ \bibnamefont
  {Djimbi}}, \bibinfo {author} {\bibfnamefont {J.}~\bibnamefont {Roques}},
  \bibinfo {author} {\bibfnamefont {H.}~\bibnamefont {Balout}}, \bibinfo
  {author} {\bibfnamefont {R.~A.}\ \bibnamefont {Ketcham}}, \bibinfo {author}
  {\bibfnamefont {E.}~\bibnamefont {Simoni}}, \bibinfo {author} {\bibfnamefont
  {R.}~\bibnamefont {Pik}}, \bibinfo {author} {\bibfnamefont {A.-M.}\
  \bibnamefont {Seydoux-Guillaume}},\ and\ \bibinfo {author} {\bibfnamefont
  {L.}~\bibnamefont {Tassan-Got}},\ }\href@noop {} {\bibfield  {journal}
  {\bibinfo  {journal} {Geochimica et Cosmochimica Acta}\ }\textbf {\bibinfo
  {volume} {268}},\ \bibinfo {pages} {348} (\bibinfo {year}
  {2020})}\BibitemShut {NoStop}%
\bibitem [{\citenamefont {Ketcham}\ \emph {et~al.}(2013)\citenamefont
  {Ketcham}, \citenamefont {Guenthner},\ and\ \citenamefont
  {Reiners}}]{ketcham2013geometric}%
  \BibitemOpen
  \bibfield  {author} {\bibinfo {author} {\bibfnamefont {R.~A.}\ \bibnamefont
  {Ketcham}}, \bibinfo {author} {\bibfnamefont {W.~R.}\ \bibnamefont
  {Guenthner}},\ and\ \bibinfo {author} {\bibfnamefont {P.~W.}\ \bibnamefont
  {Reiners}},\ }\href@noop {} {\bibfield  {journal} {\bibinfo  {journal}
  {American Mineralogist}\ }\textbf {\bibinfo {volume} {98}},\ \bibinfo {pages}
  {350} (\bibinfo {year} {2013})}\BibitemShut {NoStop}%
\bibitem [{\citenamefont {Reich}\ \emph {et~al.}(2007)\citenamefont {Reich},
  \citenamefont {Ewing}, \citenamefont {Ehlers},\ and\ \citenamefont
  {Becker}}]{reich2007low}%
  \BibitemOpen
  \bibfield  {author} {\bibinfo {author} {\bibfnamefont {M.}~\bibnamefont
  {Reich}}, \bibinfo {author} {\bibfnamefont {R.~C.}\ \bibnamefont {Ewing}},
  \bibinfo {author} {\bibfnamefont {T.~A.}\ \bibnamefont {Ehlers}},\ and\
  \bibinfo {author} {\bibfnamefont {U.}~\bibnamefont {Becker}},\ }\href@noop {}
  {\bibfield  {journal} {\bibinfo  {journal} {Geochimica et Cosmochimica Acta}\
  }\textbf {\bibinfo {volume} {71}},\ \bibinfo {pages} {3119} (\bibinfo {year}
  {2007})}\BibitemShut {NoStop}%
\bibitem [{\citenamefont {Saadoune}\ \emph {et~al.}(2009)\citenamefont
  {Saadoune}, \citenamefont {Purton},\ and\ \citenamefont
  {de~Leeuw}}]{saadoune2009}%
  \BibitemOpen
  \bibfield  {author} {\bibinfo {author} {\bibfnamefont {I.}~\bibnamefont
  {Saadoune}}, \bibinfo {author} {\bibfnamefont {J.~A.}\ \bibnamefont
  {Purton}},\ and\ \bibinfo {author} {\bibfnamefont {N.~H.}\ \bibnamefont
  {de~Leeuw}},\ }\href {https://doi.org/10.1016/j.chemgeo.2008.10.015}
  {\bibfield  {journal} {\bibinfo  {journal} {Chemical Geology}\ }\textbf
  {\bibinfo {volume} {258}},\ \bibinfo {pages} {182} (\bibinfo {year}
  {2009})}\BibitemShut {NoStop}%
\bibitem [{\citenamefont {Todorov}\ \emph {et~al.}(2006)\citenamefont
  {Todorov}, \citenamefont {Smith}, \citenamefont {Trachenko},\ and\
  \citenamefont {Dove}}]{todorov_smith_trachenko_dove_2006}%
  \BibitemOpen
  \bibfield  {author} {\bibinfo {author} {\bibfnamefont {I.~T.}\ \bibnamefont
  {Todorov}}, \bibinfo {author} {\bibfnamefont {W.}~\bibnamefont {Smith}},
  \bibinfo {author} {\bibfnamefont {K.}~\bibnamefont {Trachenko}},\ and\
  \bibinfo {author} {\bibfnamefont {M.~T.}\ \bibnamefont {Dove}},\ }\href
  {https://doi.org/10.1039/b517931a} {\bibfield  {journal} {\bibinfo  {journal}
  {Journal of Materials Chemistry}\ }\textbf {\bibinfo {volume} {16}},\
  \bibinfo {pages} {1911} (\bibinfo {year} {2006})}\BibitemShut {NoStop}%
\bibitem [{\citenamefont {Grimes}\ \emph {et~al.}(1990)\citenamefont {Grimes},
  \citenamefont {Miller},\ and\ \citenamefont {CATLOW}}]{grimes1990behaviour}%
  \BibitemOpen
  \bibfield  {author} {\bibinfo {author} {\bibfnamefont {R.~W.}\ \bibnamefont
  {Grimes}}, \bibinfo {author} {\bibfnamefont {R.~H.}\ \bibnamefont {Miller}},\
  and\ \bibinfo {author} {\bibfnamefont {C.~A.}\ \bibnamefont {CATLOW}},\
  }\href@noop {} {\bibfield  {journal} {\bibinfo  {journal} {Journal of nuclear
  materials}\ }\textbf {\bibinfo {volume} {172}},\ \bibinfo {pages} {123}
  (\bibinfo {year} {1990})}\BibitemShut {NoStop}%
\bibitem [{\citenamefont {Govers}\ \emph {et~al.}(2009)\citenamefont {Govers},
  \citenamefont {Lemehov}, \citenamefont {Hou},\ and\ \citenamefont
  {Verwerft}}]{govers2009molecular}%
  \BibitemOpen
  \bibfield  {author} {\bibinfo {author} {\bibfnamefont {K.}~\bibnamefont
  {Govers}}, \bibinfo {author} {\bibfnamefont {S.}~\bibnamefont {Lemehov}},
  \bibinfo {author} {\bibfnamefont {M.}~\bibnamefont {Hou}},\ and\ \bibinfo
  {author} {\bibfnamefont {M.}~\bibnamefont {Verwerft}},\ }\href@noop {}
  {\bibfield  {journal} {\bibinfo  {journal} {Journal of nuclear materials}\
  }\textbf {\bibinfo {volume} {395}},\ \bibinfo {pages} {131} (\bibinfo {year}
  {2009})}\BibitemShut {NoStop}%
\bibitem [{\citenamefont {Yang}\ \emph {et~al.}(2014)\citenamefont {Yang},
  \citenamefont {Zarkadoula}, \citenamefont {Dove}, \citenamefont {Todorov},
  \citenamefont {Geisler}, \citenamefont {Brazhkin},\ and\ \citenamefont
  {Trachenko}}]{yang2014solid}%
  \BibitemOpen
  \bibfield  {author} {\bibinfo {author} {\bibfnamefont {C.}~\bibnamefont
  {Yang}}, \bibinfo {author} {\bibfnamefont {E.}~\bibnamefont {Zarkadoula}},
  \bibinfo {author} {\bibfnamefont {M.}~\bibnamefont {Dove}}, \bibinfo {author}
  {\bibfnamefont {I.}~\bibnamefont {Todorov}}, \bibinfo {author} {\bibfnamefont
  {T.}~\bibnamefont {Geisler}}, \bibinfo {author} {\bibfnamefont
  {V.}~\bibnamefont {Brazhkin}},\ and\ \bibinfo {author} {\bibfnamefont
  {K.}~\bibnamefont {Trachenko}},\ }\href@noop {} {\bibfield  {journal}
  {\bibinfo  {journal} {Journal of Applied Physics}\ }\textbf {\bibinfo
  {volume} {116}},\ \bibinfo {pages} {184901} (\bibinfo {year}
  {2014})}\BibitemShut {NoStop}%
\bibitem [{\citenamefont {Saadoune}\ and\ \citenamefont
  {De~Leeuw}(2009)}]{saadoune2009computer}%
  \BibitemOpen
  \bibfield  {author} {\bibinfo {author} {\bibfnamefont {I.}~\bibnamefont
  {Saadoune}}\ and\ \bibinfo {author} {\bibfnamefont {N.~H.}\ \bibnamefont
  {De~Leeuw}},\ }\href@noop {} {\bibfield  {journal} {\bibinfo  {journal}
  {Geochimica et Cosmochimica Acta}\ }\textbf {\bibinfo {volume} {73}},\
  \bibinfo {pages} {3880} (\bibinfo {year} {2009})}\BibitemShut {NoStop}%
\bibitem [{\citenamefont {Anderson}\ \emph {et~al.}(2020)\citenamefont
  {Anderson}, \citenamefont {van Soest}, \citenamefont {Hodges},\ and\
  \citenamefont {Hanchar}}]{anderson2020helium}%
  \BibitemOpen
  \bibfield  {author} {\bibinfo {author} {\bibfnamefont {A.~J.}\ \bibnamefont
  {Anderson}}, \bibinfo {author} {\bibfnamefont {M.~C.}\ \bibnamefont {van
  Soest}}, \bibinfo {author} {\bibfnamefont {K.~V.}\ \bibnamefont {Hodges}},\
  and\ \bibinfo {author} {\bibfnamefont {J.~M.}\ \bibnamefont {Hanchar}},\
  }\href@noop {} {\bibfield  {journal} {\bibinfo  {journal} {Geochimica et
  Cosmochimica Acta}\ }\textbf {\bibinfo {volume} {274}},\ \bibinfo {pages}
  {45} (\bibinfo {year} {2020})}\BibitemShut {NoStop}%
\bibitem [{\citenamefont {Weber}(1993)}]{weber1993alpha}%
  \BibitemOpen
  \bibfield  {author} {\bibinfo {author} {\bibfnamefont {W.~J.}\ \bibnamefont
  {Weber}},\ }\href@noop {} {\bibfield  {journal} {\bibinfo  {journal} {Journal
  of the American Ceramic Society}\ }\textbf {\bibinfo {volume} {76}},\
  \bibinfo {pages} {1729} (\bibinfo {year} {1993})}\BibitemShut {NoStop}%
\bibitem [{\citenamefont {Mir}\ \emph {et~al.}(2021)\citenamefont {Mir},
  \citenamefont {Hyatt},\ and\ \citenamefont {Donnelly}}]{mir2021situ}%
  \BibitemOpen
  \bibfield  {author} {\bibinfo {author} {\bibfnamefont {A.~H.}\ \bibnamefont
  {Mir}}, \bibinfo {author} {\bibfnamefont {N.~C.}\ \bibnamefont {Hyatt}},\
  and\ \bibinfo {author} {\bibfnamefont {S.~E.}\ \bibnamefont {Donnelly}},\
  }\href@noop {} {\bibfield  {journal} {\bibinfo  {journal} {Journal of Nuclear
  Materials}\ ,\ \bibinfo {pages} {152836}} (\bibinfo {year}
  {2021})}\BibitemShut {NoStop}%
\end{thebibliography}%

\end{document}